\documentclass[sigconf,natbib=false]{acmart}
\usepackage{graphicx}
\AtBeginDocument{%
  }

\copyrightyear{2023}
\acmYear{2023}
\setcopyright{rightsretained}
\acmConference[ITiCSE 2023]{Proceedings of the 2023 Conference on Innovation and Technology in Computer Science Education V. 1}{July 8--12, 2023}{Turku, Finland}
\acmBooktitle{Proceedings of the 2023 Conference on Innovation and Technology in Computer Science Education V. 1 (ITiCSE 2023), July 8--12, 2023, Turku, Finland}\acmDOI{10.1145/3587102.3588848}
\acmISBN{979-8-4007-0138-2/23/07}

\makeatletter
\gdef\@copyrightpermission{
  \begin{minipage}{0.3\columnwidth}
   \href{https://creativecommons.org/licenses/by/4.0/}{\includegraphics[width=0.90\textwidth]{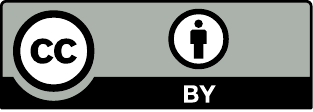}}
  \end{minipage}\hfill
  \begin{minipage}{0.7\columnwidth}
   \href{https://creativecommons.org/licenses/by/4.0/}{This work is licensed under a Creative Commons Attribution International 4.0 License.}
  \end{minipage}
  \vspace{5pt}
}
\makeatother

\RequirePackage[
  datamodel=acmdatamodel,
  style=acmnumeric,
  ]{biblatex}

\begin{document}

% \title{Towards a Success Model for Automated Programming Assessment Systems \csadd{for Formative Assessment and Feedback}}
\title{Towards a Success Model for Automated Programming Assessment Systems Used as a Formative Assessment Tool}

%%
%% The "author" command and its associated commands are used to define
%% the authors and their affiliations.
%% Of note is the shared affiliation of the first two authors, and the
%% "authornote" and "authornotemark" commands
%% used to denote shared contribution to the research.
\author{Clemens Sauerwein}
\authornote{Both authors contributed equally to this research.}
\email{Clemens.Sauerwein@uibk.ac.at}
\orcid{0009-0009-9464-5080}
\affiliation{
 \institution{University of Innsbruck}
 \streetaddress{Technikerstraße 21a}
 \city{Innsbruck}
 \country{Austria}
 \postcode{A-6020}
}

\author{Tobias Antensteiner}
\authornotemark[1]
\email{Tobias.Antensteiner@uibk.ac.at}
\orcid{0000-0001-5513-1073}
\affiliation{
 \institution{University of Innsbruck}
 \streetaddress{Technikerstraße 21a}
 \city{Innsbruck}
 \country{Austria}
 \postcode{A-6020}
}

\author{Stefan Oppl}
\email{Stefan.Oppl@donau-uni.ac.at}
\orcid{0000-0002-5495-7727}
\affiliation{
 \institution{University for Continuing Education Krems }
 \streetaddress{Dr.-Karl-Dorrek-Straße 30}
 \city{Krems}
 \country{Austria}
 \postcode{AT-3500}
}

\author{Iris Groher}
\email{Iris.Groher@jku.at}
\orcid{0000-0003-0905-6791}
\affiliation{
 \institution{Johannes Kepler University Linz}
 \streetaddress{Altenbergerstrasse 69}
 \city{Linz}
 \country{Austria}
 \postcode{AT-4040}
}

\author{Alexander Meschtscherjakov}
\email{Alexander.Meschtscherjakov@plus.ac.at}
\orcid{0000-0001-8116-4522}
\affiliation{
 \institution{Paris Lodron University of Salzburg}
 \streetaddress{Jakob-Haringer-Straße 8}
 \city{Salzburg}
 \country{Austria}
 \postcode{AT-5020}
}

\author{Philipp Zech}
\email{Philipp.Zec@uibk.ac.at}
\orcid{0000-0002-4952-4337}
\affiliation{
 \institution{University of Innsbruck}
 \streetaddress{Technikerstraße 21a}
 \city{Innsbruck}
 \country{Austria}
 \postcode{A-6020}
}

\author{Ruth Breu}
\email{Ruth.Breu@uibk.ac.at}
\orcid{0000-0001-7093-4341}
\affiliation{
 \institution{University of Innsbruck}
 \streetaddress{Technikerstraße 21a}
 \city{Innsbruck}
 \country{Austria}
 \postcode{A-6020}
}
\renewcommand{\shortauthors}{Clemens Sauerwein et al.}
%%
%% By default, the full list of authors will be used in the page
%% headers. Often, this list is too long, and will overlap
%% other information printed in the page headers. This command allows
%% the author to define a more concise list
%% of authors' names for this purpose.
%\renewcommand{\shortauthors}{Trovato et al.}

%%
%% The abstract is a short summary of the work to be presented in the
%% article.
\begin{abstract}
The assessment of source code in university education is a central and important task for lecturers of programming courses. In doing so, educators are confronted with growing numbers of students having increasingly diverse prerequisites, a shortage of tutors, and highly dynamic learning objectives. To support lecturers in meeting these challenges, the use of automated programming assessment systems (APASs), facilitating formative assessments by providing timely, objective feedback, is a promising solution. Measuring the effectiveness and success of these platforms is crucial to understanding how such platforms should be designed, implemented, and used. However, research and practice lack a common understanding of aspects influencing the success of APASs. To address these issues, we have devised a success model for APASs based on established models from information systems as well as blended learning research and conducted an online survey with $414$ students using the same APAS. In addition, we examined the role of mediators intervening between technology-, system- or self-related factors, respectively, and the users' satisfaction with APASs. Ultimately, our research has yielded a model of success comprising seven constructs influencing user satisfaction with an APAS. 
\end{abstract}

%%
%% The code below is generated by the tool at http://dl.acm.org/ccs.cfm.
%% Please copy and paste the code instead of the example below.
%%


%%
%% Keywords. The author(s) should pick words that accurately describe
%% the work being presented. Separate the keywords with commas.

%%
%% This command processes the author and affiliation and title
%% information and builds the first part of the formatted document.
\maketitle

\section{Introduction}
\label{sec:introduction}
In the course of the advancing digitization of the economy and society, programming education at universities is gaining an increasingly important role. 
While in the past programming skills played an essential role mainly in the natural sciences, their relevance is increasing in almost all fields of science. To meet this increased demand for programming education, higher education institutions offer a variety of new teaching formats such as supplementary studies or minor subjects with a strong focus on programming education. Consequently, the group of students acquiring programming skills is much larger and has heterogeneous previous knowledge as well as ways of thinking. The situation is exacerbated by a shortage of teaching staff and highly dynamic learning content and technologies~\cite{mekterovic2017}.

Therefore, solutions are needed to efficiently support teachers in managing courses and assignments for a large number of students from different backgrounds. These solutions should provide students with automatic and objective formative feedback or targeted and timely assistance with programming problems~\cite{mekterovic2017}. To meet these requirements, various solutions entered the market in recent years under the umbrella term Automated Programming Assessment System (APAS)
~\cite{keuning2018,mekterovic2020}.
According to \citeauthor{mekterovic2020} and \citeauthor{keuning2018}, APASs can motivate students, provide a useful overview of learning progress, improve the quality of teaching and student contributions, minimize the barrier to entry into programming, standardize and objectify feedback and reduce dropout rates~\cite{mekterovic2020, keuning2018}.

Consequently, recent research has examined the usability of APAS~\cite{pettit2015c}total and how they affect students' learning and lecturers' teaching experience~\cite{pettit2015b}. While current research recommends studies of the aspects that influence acceptance and satisfaction of users~\cite{sauerwein2023}, not many attempts have been made to model and evaluate the drivers of APASs' success.

To address this research challenge, we devised and validated a comprehensive success model to determine the effectiveness of APASs from a student perspective. In doing so, we drew on the works of \citeauthor{delone1992}, whose interdependent and multi-construct view of information systems is widely accepted and has been applied in a variety of fields~\cite{delone2003}. In addition, we derived certain exogenous as well as mediator constructs from \citeauthor{zhang2020}'s blended learning success model~\cite{zhang2020} to ultimately develop a model measuring the success of APASs in terms of student satisfaction~\cite{wu2010a} utilizing eight constructs.  
The remainder of the paper is structured as follows: Section~\ref{sec:related_work} discusses related work. Section~\ref{sec:applied_research_methodology} outlines the applied research methodology. Section~\ref{sec:data_analysis_and_results} presents the statistical analysis of our devised success model. Section~\ref{subsec:discussion_of_results} discusses the key findings and limitations of our research, as well as directions for future work. Finally, Section~\ref{sec:conclusion_and_outlook} concludes our research and provides an outlook on our future work. 

\section{Related Work}
\label{sec:related_work}
Recent research in higher education addresses both the understanding and support of students learning to program and of teachers providing introductory programming courses~\cite{luxton2018}. Inspired by further research shedding light on the causes of difficulties in learning programming~\cite{gomes2007}, methods and tools such as APASs~\cite{souza2016} were developed. In this context, a ``process-oriented approach in assessing programming progress''~\cite{villamor2020} was introduced, leading to an iterative learning approach to programming~\cite{pettit2015c}.

According to \citeauthor{mekterovic2017}, the use of APASs circumvents some of the hurdles associated with the manual assessment of programming tasks, such as objectively and efficiently assessing a large number of students as well as providing timely, individual and helpful feedback~\cite{mekterovic2017}. Based on a comprehensive literature review, \citeauthor{pettit2015b} have shown evidence that using APAS enhances the teacher's learning experience and increases students' learning success~\cite{pettit2015b}. Subsequent research has examined how APAS are perceived by users~\cite{restrepo2019, barra2020, rubio2014, gordillo2019}, how easy they are to use~\cite{pettit2015c}, how they affect learning and the teaching experience~\cite{pettit2015b, queiros2020}, and what users' experiences are~\cite{sauerwein2023}.

Since e-learning systems are often considered as dedicated instances of information systems (IS), APASs in particular can be evaluated by using models that have been developed and validated in the field~\cite{mayr2022}. Accordingly, \citeauthor{zhang2020} have developed a thorough blended learning success model~\cite{zhang2020} based on \citeauthor{delone1992}'s revised IS success model~\cite{delone2003} by \textit{(i)} incorporating, among others, ``individual's attitudes''~\cite{liaw2011} like student's computer self-efficacy and motivation as ``self-related factors''~\cite{zhang2020}, and \textit{(ii)} considering learning climate~\cite{wu2010a}, blended learning flexibility~\cite{zhang2020}, task-technology fit~\cite{goodhue1995} as meditators influencing student's satisfaction~\cite{sun2008} and intention.

However, recent research has focused primarily on APASs design, implementation, and use, thus, leaving a few papers to conduct empirical studies addressing drivers shaping student's APASs satisfaction. While success models exist in other fields such as IS and e- or blended learning~\cite{wu2010b, zhu2012}, a model for gauging the success of APASs is largely lacking~\cite{sauerwein2023}. Moreover, the role of mediator constructs intervening between technology, system or self-related factors and user satisfaction with APASs has not been examined.

\section{Applied Research Methodology}
\label{sec:applied_research_methodology}
To investigate the drivers of APASs success with respect to student satisfaction, we \textit{(i)} defined a research model (cf. Section~\ref{subsec:research_model}) based on \citeauthor{delone2003}'s and \citeauthor{zhang2020}'s works \cite{delone2003, zhang2020}, \textit{(ii)} developed a research instrument by adopting approved measurement items and utilizing an online questionnaire (cf. Section~\ref{subsec:research_instrument}), and \textit{(iii)} collected data from a total of $414$ students (cf. Section~\ref{subsec:data_collection}).
\subsection{Research Model}
\label{subsec:research_model}
Our devised success model for APASs builds on the refined \citeauthor{delone2003} model of IS success~\cite{delone2003} as well as \citeauthor{zhang2020}'s blended learning success model~\cite{zhang2020} and consists (cf. Figure~\ref{fig:bootstrapped_pls_model}) of the following four exogenous constructs:
\begin{itemize}
    \item[\texttt{MOT}] (Motivation) the APAS's incentives, encouraging the user to engage in learning with the system~\cite{zhang2020},
    % the user's desire to learn using the APAS
    \item[\texttt{INQ}] (Information Quality) the desirable quality characteristics of the APAS's output or content~\cite{delone2003},
    \item[\texttt{SYQ}] (System Quality) the desirable quality characteristics of the APAS itself~\cite{delone2003}, and
    \item[\texttt{SEQ}] (Service Quality) the quality of the support provided by the APAS's provider~\cite{delone2003};
\end{itemize}
the following three mediator constructs:
\begin{itemize}
    \item[\texttt{LEC}] (Learning Climate) the user's enjoyment and pleasure when using the platform, as well as the platform's ability to motivate learning~\cite{zhang2020},
    \item[\texttt{BLF}] (Blended Learning Flexibility) the degree the APAS offers the freedom to learn at one's own pace, with effective preparation and feedback~\cite{zhang2020}, and
    \item[\texttt{TTF}] (Task-Technology Fit) the degree to which the services of an APAS meet the user's learning goals and requirements~\cite{zhang2020};
\end{itemize}
as well as the endogenous construct:
\begin{itemize}
    \item[\texttt{SAT}] (Satisfaction) the users' level of satisfaction with the APAS~\cite{delone2003}.
\end{itemize}

Consequently, following \citeauthor{zhang2020}'s approach, we \textit{(i)} adopted all endogenous constructs of \citeauthor{delone1992}'s revised IS success model~\cite{delone2003}, i.e., \texttt{INQ}, \texttt{SYQ}, and \texttt{SEQ}, \textit{(ii)} considered \texttt{MOT} as an additional independent, ``self-related factor''~\cite{zhang2020}, and \textit{(iii)} introduced the APAS-related blended learning success model constructs \texttt{LEC}~\cite{wu2010a}, \texttt{BLF}~\cite{zhang2020}, and \texttt{TTF}~\cite{goodhue1995},  as meditators potentially influencing the by \citeauthor{delone1992} proposed endogenous latent variable \texttt{SAT}~\cite{sun2008}, being explained in our success model for APASs. Therefore, in accordance with~\citeauthor{rungtusanatham2014} transmittal approach~\cite{rungtusanatham2014, memon2018} and based on \citeauthor{delone1992} as well as \citeauthor{zhang2020}, we hypothesize, each of the aforementioned mediating effects, i.e., \texttt{MOT}, \texttt{INQ}, \texttt{SYQ}, and \texttt{SEQ}, respectively, have an indirect effect on \texttt{SAT} through \texttt{LEC}, \texttt{BLF}, and \texttt{TTF}.

To empirically test our theoretical model, we converted it into a structural equation model (cf. Section~\ref{subsec:model_evaluation}), as depicted in Figure~\ref{fig:bootstrapped_pls_model}, in which, among others, the hypotheses to be further investigated are represented by arrows, i.e., directed paths, connecting the respective constructs, denoted by ellipses.

\subsection{Research Instrument}
\label{subsec:research_instrument}
Our primary research tool was a self-administered online questionnaire in LimeSurvey. In order to statistically test the constructs and their relationships, we defined corresponding questions for each construct. In doing so, we followed the works of~\cite{zhang2020, delone2003} to formulate appropriate survey questions. This approach resulted in the following question items:
\begin{itemize}
    \small{\item[$\texttt{MOT}_{1}$] The APAS motivates me to work on the tasks and the feedback.}
    \small{\item[$\texttt{MOT}_{2}$] Working with the APAS increases my MOT to solve tasks.}
    \small{\item[$\texttt{MOT}_{3}$] The APAS increases my MOT when learning the course content.}
    \small{\item[$\texttt{INQ}_{1}$] The information provided by the APAS is relevant to my learning.}
    \small{\item[$\texttt{INQ}_{2}$] The information provided by the APAS is easy to understand.}
    \small{\item[$\texttt{INQ}_{3}$] The information provided by the APAS is very good.}
    \small{\item[$\texttt{INQ}_{4}$] The information provided by the APAS is up to date.}
    \small{\item[$\texttt{INQ}_{5}$] The information provided by the APAS is complete.}
    \small{\item[$\texttt{INQ}_{6}$] The information provided by the APAS is accurate.}
    \small{\item[$\texttt{SYQ}_{1}$] The APAS offers flexibility in terms of learning time and place.}
    \small{\item[$\texttt{SYQ}_{2}$]  The APAS offers interactive feedback on the exercises.}
    \small{\item[$\texttt{SYQ}_{3}$]  The APAS offers sufficient functions for my learning process.}
    \small{\item[$\texttt{SYQ}_{4}$]  In general, the APAS is reliable.}
    \small{\item[$\texttt{SYQ}_{5}$]  In general, the response time of the APAS is reasonable.}
    \small{\item[$\texttt{SEQ}_{1}$] The help and support of the APAS is fast.}
    \small{\item[$\texttt{SEQ}_{2}$] The help and support of the APAS is reliable.}
    \small{\item[$\texttt{SEQ}_{3}$] The help and support of the APAS is accessible.}
    \small{\item[$\texttt{SEQ}_{4}$] The help and support of the APAS is practical.}
    \small{\item[$\texttt{SEQ}_{5}$] The help and support provided by the APAS is satisfactory overall.}
    \small{\item[$\texttt{LEC}_{1}$] I like to use the APAS to support my learning process.}
    \small{\item[$\texttt{LEC}_{2}$] Working with the APAS is fun.}
    \small{\item[$\texttt{LEC}_{3}$] Working with the APAS is very pleasant.}
    \small{\item[$\texttt{LEC}_{4}$] The LEC of the APAS was able to motivate my spontaneous learning.}
    \small{\item[$\texttt{BLF}_{1}$] The APAS gives me more flexibility to learn in this course.}
    \small{\item[$\texttt{BLF}_{2}$] The APAS allows me to complete the exercises at my own pace.}
    \small{\item[$\texttt{BLF}_{3}$] The APAS allows me to prepare more effectively for each lesson.}
    \small{\item[$\texttt{TTF}_{1}$] The services offered by the APAS meet my learning needs.}
    \small{\item[$\texttt{TTF}_{2}$] The services of the APAS are compatible with my learning needs.}
    \small{\item[$\texttt{TTF}_{3}$] The services of the APAS fit my learning needs.}
    \small{\item[$\texttt{TTF}_{4}$] My learning goals and needs are met through the use of the APAS.}
    \small{\item[$\texttt{SAT}_{1}$] Overall, I am satisfied with the APAS.}
    \small{\item[$\texttt{SAT}_{2}$] Overall, I find the APAS enjoyable.}
    \small{\item[$\texttt{SAT}_{3}$] Overall, the APAS meets my learning needs.}
\end{itemize}

All question items were given a five-point Likert scale with $1$ for ``strongly disagree'', $3$ for ``neither agree nor disagree'' and $5$ for ``strongly agree''. In addition, demographic information such as gender, (academic) degree, number of semesters studied, and known programming languages were collected through the survey in order to characterize the sample.

\subsection{Data Collection}
\label{subsec:data_collection}
The research model (cf. Section~\ref{subsec:research_model}) was validated in the form of a survey (cf. Section~\ref{subsec:research_instrument}) in introductory programming courses at the end of the semester in February 2021 and 2022. For these introductory programming courses, we used the APAS ArTeMis~\cite{krusche2018} as it provides all the basic functions, such as course, exercise and test management, learning analytics, just-in-time teaching, and automated feedback features. In addition, the same adapted blended learning method was used in each course and an introduction to programming was taught. To reinforce the learning content at home, a weekly exercise sheet was made available via ArTeMis. This exercise sheet consisted of several programming tasks with stored test cases and formative feedback in ArTeMis.  The students had to work on these exercise sheets independently and were supported by appropriate feedback from ArTeMis. After successfully completing the exercise sheets, the students had to participate in a weekly video conference to discuss the results and receive targeted assistance from the lecturers. Due to the pandemic situation, video conferences had to be held remotely online instead of face-to-face meetings at the university. 

A total of $414$ students -- more than $75\%$ were seeking a Bachelor's degree, about $20\%$ were enrolled in a Master's program and $5\%$ hold a Master's degree -- participated in our survey (cf. Section~\ref{subsec:research_instrument}). After removing a total of $46$ partially completed questionnaires, i.e., responses containing missing values (MVs) in at least one descriptive statistics data's or all research model item's columns (cf. Section~\ref{subsec:research_model}), $368$ records remained for statistical analysis. Thus, our dataset contains $5.57\%$ item non-response MVs, which we can attribute to partially completed questionnaires. However, these missing values do not affect our data analysis, as the employed estimation method (cf. Section~\ref{sec:data_analysis_and_results}) provides unbiased results up to a MV percentage of $8.89\%$~\cite{grimm2020}.

Of the overall $98$ female, $262$ male students, and $8$ with no indication of gender who participated in our survey, $271$ were in their first semester of study or had completed it recently. Consequently, about $40\%$, i.e., in total $156$, were not familiar with any well-known (general-purpose) programming language such as C, Java or Python. Taken together, $132$ of the students surveyed stated to know Python, less than one quarter, i.e., overall $88$, have heard of Java, $77$ were acquainted with C, and $58$ were familiar with C++ before enrolling in the respective (introductory) course.

\section{Data Analysis and Results}
\label{sec:data_analysis_and_results}
In order to conduct a systematic mediation analysis (cf. Section~\ref{subsec:mediation_analysis}), which builds on our theoretically established success model for APASs and hypothesized relationships (cf. Section~\ref{subsec:research_model}),
we estimate and assess the corresponding structural equation model (cf. Section~\ref{subsec:model_evaluation})~\cite{hair2021b}.

\begin{figure*}
\centering
\includegraphics[width=\textwidth]{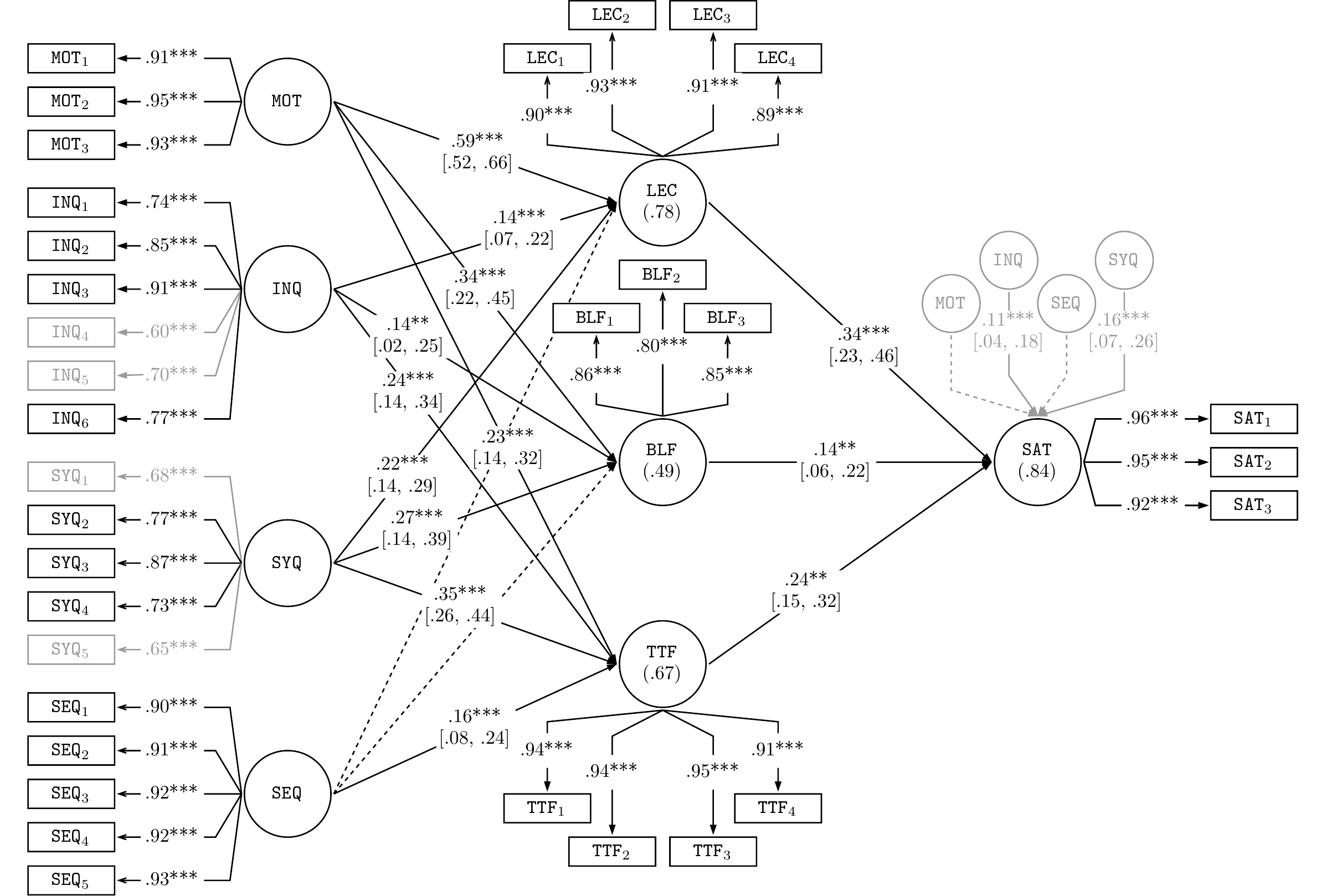}
\Description[Bootstrapped PLS model using $10\,000$ bootstrap samples.]{Bootstrapped PLS model ($\texttt{nboot} = 10\,000$).}
\caption[Bootstrapped PLS model ($\texttt{nboot} = 10\,000$).]{Bootstrapped PLS model using $10\,000$ bootstrap samples.}
\footnotesize{Coefficients of non-significant paths, i.e., dotted lines, are not shown. Items marked in gray refer to eliminated indicators. Values in braces denote the $R^{2}$ value of the respective (endogenous) construct, while value pairs in brackets indicate the $95\%$ confidence interval of the corresponding significant path. Stars refer to the significance level of the respective factor loading or path coefficient, where three stars (``***'') denotes $p \in [0, .001]$ and two stars (``**'') denotes $p \in (.001, 0.01]$. Furthermore, direct effects used for mediation analysis are highlighted in gray; thus, both, the ordinary and the gray plotted version of a respective exogenous construct refer to the same latent variable.}
\label{fig:bootstrapped_pls_model}
\end{figure*}

\subsection{Model Evaluation}
\label{subsec:model_evaluation} 
Structural equation modeling (SEM) is a useful technique for evaluating complex theoretical relationships, especially between multiple latent variables, i.e., exogenous and endogenous constructs, in particular when conducting research, such as our study, in social science~\cite{hair2022}. Two fundamental SEM methods, covariance-based SEM (CB-SEM) and PLS-SEM have been proposed and dominate SEM in practice~\cite{hair2021a}. However, as in our case, PLS-SEM is particularly useful when the goal of the structural model is to predict and explain the intended outcomes as ``obtained by the in-sample and out-of-sample metrics''~\cite{latan2017}. Accordingly, we used the statistical computing language R version \textit{4.2.} and leveraged the \texttt{SEMinR} package~\cite{ray2021} version \textit{2.3.} to test our research hypotheses (cf. Section~\ref{subsec:research_model}) employing a PLS-SEM analysis~\cite{ray2021, hair2021a}.

\subsubsection{Measurement Model Evaluation}
\label{subsubsec:measurement_model_evaluation}
Starting with the reliability and validity evaluation of our reflective measurement by following \citeauthor{hair2021a}, we \textit{(i)} assess the indicator reliability using the respective indicator loadings, \textit{(ii)} examine the internal consistency reliability leveraging composite reliability ($\rho_{c}$), Cronbach's alpha ($\rho_{T}$), as well as the reliability coefficient ($\rho_{A}$), \textit{(iii)} assess the convergent validity based on the average variance extracted (AVE), and \textit{(iv)} check the discriminant validity through the corresponding Heterotrait-Monotrait (HTMT) ratios~\cite{hair2019a}.

Regarding \textit{(i)}, all indicator loadings, except $\texttt{INQ}_{4}$, $\texttt{INQ}_{5}$, $\texttt{SYQ}_{1}$, and $\texttt{SYQ}_{5}$ (cf. Sections~\ref{subsec:research_instrument}), of the eight (reflectively) measured constructs \texttt{MOT}, \texttt{INQ}, \texttt{SYQ}, \texttt{SEQ}, \texttt{LEC}, \texttt{BLF}, \texttt{TTF}, and \texttt{SAT} (cf. Sections~\ref{subsec:research_model}) are statistically significant at $\text{CI}_{\alpha} = .05$ or below as well as higher than the threshold value of $.708$~\cite{hair2021b}, which suggests sufficient levels of indicator reliability~\cite{hair2021a}. As according to \citeauthor{hair2021a}, ``indicators with loadings between $.40$ and $.708$ should be considered for removal only when deleting the indicator leads to an increase in the internal consistency reliability or convergent validity'' (cf. \textit{(ii)} and \textit{(iii)}), we eliminate  $\texttt{INQ}_{4}$, $\texttt{INQ}_{5}$, $\texttt{SYQ}_{1}$, as well as $\texttt{SYQ}_{5}$ (cf. Figure~\ref{fig:bootstrapped_pls_model}), as the deletion increases the respective AVE values (cf. \textit{(iii)}), but does not affect the respective internal consistency reliability.

As for \textit{(ii)}, all eight (reflectively) measured constructs show high levels of internal consistency~\cite{hair2021b}, as $\rho_{c}$, $\rho_{T}$ and $\rho_{A}$ are all above $.70$ and less than $.95$~\cite{hair2021a}.
Moreover, concerning \textit{(iii)}, the respective AVE values are all above the cut-off value of $.50$~\cite{hair2022}, implying the measures of the eight (reflectively) measured constructs \texttt{MOT}, \texttt{INQ}, \texttt{SYQ}, \texttt{SEQ}, \texttt{LEC}, \texttt{BLF}, \texttt{TTF}, and \texttt{SAT} (cf. Sections~\ref{subsec:research_model}) have high levels of convergent validity~\cite{hair2021a}.

With respect to \textit{(iv)}, all HTMT ratio values are below the liberal cut-off value of $.90$~\cite{hair2022} for conceptually similar constructs such as \texttt{MOT} and \texttt{LEC} or \texttt{SYQ} and \texttt{TTF} and lower than the conservative cut-off value of $.85$~\cite{hair2021a} for conceptually distinct constructs such as \texttt{SEQ} and \texttt{BLF}, indicating discriminant validity between the eight (reflectively) measured constructs. Furthermore, as recommended by \citeauthor{hair2021a}, the upper bound of the $95\%$ bootstrap confidence intervals show that the HTMT ratio values are significantly lower than the liberal threshold value of $.90$.

\subsubsection{Structural Model Evaluation}
\label{subsubsec:structural_model_evaluation}
Having established the reliability and validity of the constructs, we examine the structural component of our devised success model for APASs ``based on its ability to predict the outcomes.''~\cite{hair2022} Following ~\citeauthor{hair2022}'s recommendations, we \textit{(i)} examine the structural model for collinearity issues based on the variance inflation factor (VIF), \textit{(ii)} assess the significance and relevance of the structural model relationships, i.e, the path coefficients, using bootstrapping, \textit{(iii)} evaluate the structural model's explanatory power by the coefficient of determination ($R^{2}$) and the effect size ($f^{2}$), and \textit{(iv)} examine its out-of-sample predictive power using the PLSpredict~\cite{shmueli2019} method.

As for \textit{(i)}, the VIF results indicate, our model does not show signs of collinearity among the eight constructs~\cite{hair2021a}, as the highest VIF value are close to the conservative threshold of $3.3$~\cite{hair2019b, hair2022}.

Regarding \textit{(ii)}, we assess the significance as well as the relevance of the structural model paths by bootstrapping the sampling distribution~\cite{preacher2008, zhao2010, nitzl2016} with $10\,000$~\cite{hair2021a} resamples to test the structural model's relationships coefficients for statistical significance at $\text{CI}_{\alpha} = .05$, as summarized in Figure~\ref{fig:bootstrapped_pls_model}. 

With respect to \textit{(iii)}, the $R^{2}$ values (cf. Figure~\ref{fig:bootstrapped_pls_model}) of the endogenous constructs \texttt{LEC} and \texttt{SAT} as well as of the endogenous constructs \texttt{BLF} and \texttt{TTF} are substantial as well as moderate~\cite{hair2021a}, respectively, indicating a good in-sample predictive power~\cite{hair2021b} of our success model for APASs. Moreover, the $f^{2}$ values for each of the structural model paths follow the same rank order as corresponding  path coefficient magnitude~\cite{hair2021a}.

Concerning \textit{(iv)}, our analysis shows that our bootstrapped PLS model has lower out-of-sample predictive error compared to the naïve linear regression model benchmark for the minority of the indicators of the outcome variable substantially explained by the predictor construct \texttt{SAT}, indicating low predictive power~\cite{hair2021a} and an acceptable claim of external validity for similar contexts~\cite{shmueli2016}.

\subsection{Mediation Analysis}
\label{subsec:mediation_analysis}
As described by \citeauthor{hair2021a} ``mediation occurs when a construct, referred to as mediator construct, intervenes between two other related constructs.''~\cite[p.~140]{hair2021a} Thus, for example, a change in the exogenous construct $\texttt{MOT}$ may cause a change in the theoretically considered~\cite{memon2018} mediator constructs $\texttt{LEC}$, $\texttt{BLF}$, and $\texttt{TTF}$~\cite{zhang2020, delone2003} (cf. Section~\ref{subsec:research_model}), respectively, possibly leading to a change in the endogenous construct $\texttt{SAT}$ (cf. Figure~\ref{fig:bootstrapped_pls_model}), each thereby representing an indirect mediating effect~\cite{hair2021a, memon2018}.

Since all quality criteria of the measurement and
structural model are met (cf. Section~\ref{subsubsec:measurement_model_evaluation} and Section~\ref{subsubsec:structural_model_evaluation}, respectively) and, in particular, as recommended by \citeauthor{hair2021a}, all ``reflectively measured mediator constructs exhibit a high level of reliability''~\cite[p.~141]{hair2021a}, high construct collinearity is not present and the construct’s discriminant validity are below the liberal cut-off value of $.90$~\cite{hair2021a, hair2022}, our reflective measurement model as well as the structural model are sound to perform mediation analysis.

Following \citeauthor{carrion2017}'s guidelines for multiple mediation analysis in PLS-SEM~\cite{carrion2017} based on \citeauthor{nitzl2016}~\cite{nitzl2016}, \citeauthor{zhao2010}~\cite{zhao2010} as well as \citeauthor{baron1986}~\cite{baron1986}, we first test for significance of the indirect effects~\cite{carrion2017, hair2021a} from the exogenous constructs $\texttt{MOT}$, $\texttt{INQ}$, $\texttt{SYQ}$ and $\texttt{SEQ}$ via $\texttt{LEC}$, $\texttt{BLF}$, and $\texttt{TTF}$ to endogenous constructs $\texttt{SAT}$, respectively, (cf. Figure~\ref{fig:bootstrapped_pls_model}) by bootstrapping the sampling distribution~\cite{preacher2008, zhao2010, nitzl2016}. Since, according to \citeauthor{carrion2017}, ``if the significance of each indirect effect cannot be established, there is no mediating effect''~\cite[p.~183]{carrion2017}, we infer that each of the three respective indirect effects~\cite{nitzl2016} as well as the resulting total indirect effects (TIE)~\cite{carrion2017} $\text{TIE}_{\texttt{MOT}\xrightarrow{\texttt{LEC}, \texttt{BLF}, \texttt{TTF}}\texttt{SAT}} \approx .20 + .04 + .06 = .30$, $\text{TIE}_{\texttt{INQ}\xrightarrow{\texttt{LEC}, \texttt{BLF}, \texttt{TTF}}\texttt{SAT}} \approx .05 + .02 + .05 = .12$, and $\text{TIE}_{\texttt{SYQ}\xrightarrow{\texttt{LEC}, \texttt{BLF}, \texttt{TTF}}\texttt{SAT}} \approx .07 + .04 + .08 = .19$ are significant at $\text{CI}_{\alpha} = .05$ by inspecting the bootstrap confidence intervals (CIs)~\cite{hair2021a}, respectively.

In accordance with \citeauthor{carrion2017} mediation analysis procedure~\cite{carrion2017, nitzl2016}, we secondly assess whether the relevant direct effects (DE)~\cite{zhao2010, hair2021a} (cf. Figure~\ref{fig:bootstrapped_pls_model}) are significant for each of the three mediation effects previously determined. As $\text{DE}_{\texttt{MOT}\rightarrow\texttt{SAT}}$ is not significant and $\text{DE}_{\texttt{INQ}\rightarrow\texttt{SAT}} \approx .11$ as well as $\text{DE}_{\texttt{SYQ}\rightarrow\texttt{SAT}} \approx .16$ are significant at $\text{CI}_{\alpha} = .05$, respectively, we conclude that \textit{(i)} the relationship between exogenous construct $\texttt{MOT}$ and endogenous construct $\texttt{SAT}$ is fully mediated by $\texttt{LEC}$, $\texttt{BLF}$, and $\texttt{TTF}$, and \textit{(ii)} the exogenous constructs $\texttt{LEC}$, $\texttt{BLF}$, and $\texttt{TTF}$ partially mediate the effect of $\texttt{INQ}$ respectively $\texttt{SYQ}$ on the endogenous construct $\texttt{SAT}$. Moreover, both variance accounted for (VAF) values, $\text{VAF}_{\texttt{INQ}\xrightarrow{\texttt{LEC}, \texttt{BLF}, \texttt{TTF}}\texttt{SAT}} \approx 53\%$ and $\text{VAF}_{\texttt{SYQ}\xrightarrow{\texttt{LEC}, \texttt{BLF}, \texttt{TTF}}\texttt{SAT}} \approx 54\%$, lie in the interval $[20\%, 80\%]$~\cite{hair2021a}, thus providing us with an additional argument to imply partial mediation for the latter two complementary mediations~\cite{nitzl2016, carrion2017}.

\section{Discussion of Results}
\label{subsec:discussion_of_results}
While our data analysis has shown that the model developed is valid, we cannot support all the hypotheses raised (cf. Section~\ref{subsec:research_model}). Therefore, in the following, we discuss the key findings (cf. Section~\ref{subsec:key_findings}) and outline possible limitations of the present research as well as directions for future work (cf. Section~\ref{subsec:limit}).

\subsection{Key Findings}
\label{subsec:key_findings}
As shown in Figure~\ref{fig:bootstrapped_pls_model}, most of the model's hypotheses are supported by our study. Only the paths from \texttt{SEQ} to \texttt{LEC} and from \texttt{SEQ} to \texttt{BLF} are not statistically significant and, thus, the corresponding hypotheses are not supported.

Our investigations showed that \texttt{MOT} significantly influences \texttt{LEC}, \texttt{BLF} and \texttt{TTF}. In other words, if students are motivated by the platform to work on their assignments, this has a positive effect on \texttt{LEC}, \texttt{BLF} and \texttt{TTF} of the platform. Accordingly, \texttt{MOT} of students to use the platform should be kept high by providing a user-friendly design~\cite{sun2008}. This could be increased, for example, through targeted support for editing tasks or additional functions such as a fully-fledged integrated code editor~\cite{sauerwein2023}.

The results show that \texttt{INQ} also significantly influence \texttt{LEC}, \texttt{BLF} and \texttt{TTF}. This means that \texttt{INQ} is a critical success factor, along with the student's \texttt{MOT} to use the platform. Therefore, it is clear that the information provided by the platform, in the form of exercise sheets and feedback, should be taken seriously and utilized to its full potential. Related work showed that the preparation of suitable tasks and corresponding feedback for the students is a major challenge for lecturers~\cite{sauerwein2023}. To ensure \texttt{INQ} and the success of the platform, it is essential that teachers create assignments and test cases with the appropriate feedback. Therefore, an APAS should offer functions that are capable of aiding lecturers in meeting these challenges~\cite{mekterovic2020}.

As discussed, \texttt{LEC}, \texttt{BLF} and \texttt{TTF} are significantly impacted by \texttt{SYQ}. \texttt{SYQ} refers to the reliability, flexibility, interactive feedback and sufficient functions offered by the APAS to support the learning process. Recent research has revealed that while students using an APAS primarily make use of the core functionalities such as exercise submission and feedback, other services, such as support functions, are used only marginally~\cite{sauerwein2023}. Therefore, it is not surprising that \texttt{SEQ} only affects the \texttt{TTF}, thus playing a subordinate role. In contrast to \texttt{SYQ}, \texttt{SEQ} refers to the help and support provided by the APAS. In order to fulfil \texttt{SEQ} it implies that the APAS's service functionalities should be fast, reliable, accessible, practical and overall satisfactory

Consistent with the hypothesis made by \citeauthor{zhang2020}~\cite{zhang2020} as well as in-depth research on blended learning~\cite{dang2020}, the students' perceived \texttt{LEC}~\cite{wu2010a}, \texttt{BLF}~\cite{zhang2020}, and \texttt{TTF}~\cite{goodhue1995} significantly influences their \texttt{SAT} with the APAS. Accordingly, this points to the importance \textit{(i)} of the user's enjoyment and pleasure when using the APAS, \textit{(ii)} ``offering the desired level of flexibility in students' learning process''~\cite{zhang2020}, and \textit{(iii)} the degree to which the services of an APAS meet the user's learning needs, goals as well as requirements.

Mediation analysis (cf. Section~\ref{subsec:mediation_analysis}) reveals one ``indirect-only mediation''~\cite{hair2022} and two ``complementary mediations''~\cite{hair2022} as the three hypothesized mediator constructs \texttt{LEC}, \texttt{BLF}, and \texttt{TTF} fully mediate the relationship between \texttt{MOT} and \texttt{SAT} as well as partial mediate the relationship between \texttt{INQ} respectively \texttt{SYQ} and \texttt{SAT} (cf. Section~\ref{subsec:mediation_analysis}). Furthermore, since neither \texttt{SEQ}'s direct nor total indirect effect on \texttt{SAT} is significant, the mediation analysis shows one ``no-effect non-mediation''~\cite{hair2022} by not supporting the corresponding hypothesized mediation effect.

As \texttt{MOT} direct effect on \texttt{SAT} is not significant, the important impact of the mediating effect of \texttt{LEC}, \texttt{BLF}, and \texttt{TTF} emphasis the vital role of the students' perceptions of the learning climate, task-technology fit, and blended learning flexibility in positively influencing individual’s attitudes when using the APAS~\cite{sun2008}. Moreover, the \texttt{LEC}'s, \texttt{BLF}'s, and \texttt{TTF}'s partial mediation of the effect of $\texttt{INQ}$ respectively $\texttt{SYQ}$ on $\texttt{SAT}$ entails that factors shaping the quality characteristics of the APAS, its output and its content, i.e., APAS's technology dimension related drivers~\cite{dang2020}, not only have a significant direct effect on the APAS success~\cite{delone2003, zhang2020}, but also are positively influenced in their impact by \textit{(i)} students' overall delight and happiness in using the APAS~\cite{wu2010a}, \textit{(ii)} the degree to which the APAS provide students with self-paced learning opportunities~\cite{zhang2020}, and \textit{(iii)} the level to which an APAS's services meet students' learning concerns, objectives, and demands~\cite{lin2012}.

Finally, our research has shown which aspects should be taken into account to increase users' \texttt{SAT} with an APAS and thus ensure its successful use. This statement is reinforced by~\citeauthor{lin2008}, as the concept of \texttt{SAT} is the ultimate goal of any product and service~\cite{lin2008}. Moreover, recent research has shown~\cite{diep2017} that \texttt{SAT} is the most accepted measure of learning environments' quality and effectiveness.

\subsection{Limitations \& Directions for Future Work}
\label{subsec:limit}
The research at hand might be limited by a \textit{(i)} selection bias of participants, \textit{(ii)} the used APAS, \textit{(iii)} a large number of constructs at an abstract level, and \textit{(iv)} the used teaching method.

Regarding \textit{(i)}, we asked the students of the respective introductory courses to voluntarily participate in our survey. Accordingly, only students in computer science-related studies participated in our study. Due to this, a high computer affinity of the study participants could be assumed.

The limitation \textit{(ii)} can be invalidated to a certain extent, as a systematic analysis of the functional scope of different APASs has shown that the basic functions offered are similar in all systems~\cite{sauerwein2023}. Furthermore, the aim of our work was to identify the overarching success factors of APASs, rather than assessing individual functions in detail. However, future work is required to investigate whether our devised success model can be generalised to other APASs in order to increase the representativeness of our findings.

Concerning \textit{(iii)}, our success model consists of a large number of constructs at an abstract level (cf. Section~\ref{subsec:research_model}). This was necessary to make the different constructs comparable on a uniform level of abstraction and analyze their relationship. For this reason, some constructs, such as \texttt{MOT},\texttt{INQ}, \texttt{SYQ}, or \texttt{TTF}, were only considered superficially and should be investigated in more detail in future work.  

Due to the COVID-19 pandemic, the APAS was used in a blended learning format adapted to distance learning, resulting in students' face-to-face meetings with instructors being conducted via videoconferencing. However, limitation \textit{iv} could not be avoided due to the pandemic.

\section{Conclusion \& Outlook}
\label{sec:conclusion_and_outlook}
In this paper, by drawing on the work of \citeauthor{delone1992} and \citeauthor{zhang2020}, we introduced a success model for APASs consisting of four exogenous constructs -- \texttt{MOT}, \texttt{INQ}, \texttt{SYQ} and \texttt{SEQ} -- and three mediator constructs -- \texttt{LEC}, \texttt{BLF}, and \texttt{TTF} --, influenceing the students' \texttt{SAT} with the APAS.
We estimated the constructed model using PLS-SEM~\cite{hair2021b} based on indicator data from two online surveys among a total of $414$ students who attended introductory programming courses using the same APAS.
Our investigations have shown that each of the exogenous constructs, with the exception of \texttt{SEQ}, has a significant indirect effect on \texttt{SAT} via the hypothesized mediating constructs \texttt{LEC}, \texttt{BLF}, and \texttt{TTF}. These results emphasize \textit{(i)} the crucial role of the user's enjoyment using the APAS, \textit{(ii)} the platform's functionalities in fostering learning flexibility, and \textit{(iii)} the extent to which the APAS's services meet the user's learning needs. Furthermore, our findings should enable practitioners to better understand the key aspects influencing the success of APASs. As part of our future work, it would be interesting to investigate whether the model can also be applied to other APASs and to empirically examine important constructs, e.g., \texttt{MOT} or \texttt{TTF}, in more detail.

\printbibliography
\end{document}